\documentclass{ws-ijmpe}
\usepackage[super,compress]{cite}

\begin{document}

\markboth{Cong Pan, Kaiyuan Zhang, Shuangquan Zhang}{Multipole expansion of densities in the deformed 
relativistic Hartree-Bogoliubov theory in continuum}

\catchline{}{}{}{}{}

\title{Multipole expansion of densities in the deformed relativistic Hartree-Bogoliubov theory in continuum}

\author{Cong Pan}

\address{State Key Laboratory of Nuclear Physics and Technology, School of Physics, Peking University, Beijing 100871, China}

\author{Kaiyuan Zhang}

\address{State Key Laboratory of Nuclear Physics and Technology, School of Physics, Peking University, Beijing 100871, China}

\author{Shuangquan Zhang}

\address{State Key Laboratory of Nuclear Physics and Technology, School of Physics, Peking University, Beijing 100871, China\\
sqzhang@pku.edu.cn}

\maketitle

\begin{history}
\received{Day Month Year}
\revised{Day Month Year}
\end{history}

\begin{abstract}

The deformed relativistic Hartree-Bogoliubov theory in continuum (DRHBc) has been proved one of 
the best models to probe the exotic structures in deformed nuclei. In DRHBc, the potentials and 
densities are expressed in terms of the multipole expansion with Legendre polynomials, the dependence 
on which has only been touched for light nuclei so far. In this paper, taking a light nucleus $^{20}$Ne 
and a heavy nucleus $^{242}$U as examples, we investigated the dependence on the multipole expansion 
of the potentials and densities in DRHBc. It is shown that the total energy converges well with the 
expansion truncation both in the absence of and presence of the pairing correlation, either in the 
ground state or at a constrained quadrupole deformation. It is found that to reach a same accuracy 
of the total energy, even to a same relative accuracy by percent, a larger truncation is required 
by a heavy nucleus than a light one. The dependence of the total energy on the truncation increases 
with deformation. By decompositions of the neutron density distribution, it is shown that a higher-order 
component has a smaller contribution. With the increase of deformation, the high-order components get 
larger, while at the same deformation, the high-order components of a heavy nucleus play a more important 
role than that of a light one.

\end{abstract}


\ccode{PACS numbers: 21.60.Jz, 21.10.Gv, 21.10.Dr}


\section{Introduction}

The study of exotic nuclei has been one of the frontiers in both experimental and theoretical nuclear physics research
\cite{Tanihata1995PPNP,Sorlin2008PPNP,Alkhazov2011IJMPE,Pfutzner2012RMP,Tanihata2013PPNP,Savran2013PPNP,Nakamura2017PPNP,
Brown2001PPNP,Vretenar2005PR,Meng2006PPNP,Meng2015JPG,Meng2016book,Zhou2016PoS,Qi2019PPNP}. Although more and more exotic 
nuclei are experimentally produced with the development of radioactive ion beam facilities around the world 
\cite{nndc,Thoennessen2004RPP,Thoennessen2013RPP,Wang2016AME,Nakamura2017PPNP}, there are still a large amount of predicted 
nuclei far beyond the experimental capability \cite{Erler2012Nat,Xia2018ADNDT}. Therefore, to predict accurately the 
unknown properties of these nuclei, reliable microscopic theoretical methods are welcome.

The density functional theory (DFT) and its covariant version (CDFT) have provided successful descriptions 
for many nuclear phenomena, and have attracted wide attention in nuclear physics research
\cite{Bender2003RMP,Ring1996PPNP,Vretenar2005PR,Meng2006PPNP,Niksic2011PPNP,Meng2015JPG,Meng2016book,Liang2015PR,Zhao2018IJMPE}.
Based on the CDFT, the relativistic continuum Hartree-Bogoliubov (RCHB) theory considers the pairing 
correlation and continuum in a unified and self-consistent way \cite{Meng1996PRL,Meng1998NPA}, and has 
successfully provided the descriptions for many exotic nuclear phenomena \cite{Meng1996PRL,Meng1998NPA,
Meng1998PRL,Meng1998PLB,Meng1998PRC,Meng1999PRC,Meng2002PRC,Meng2002PLB,Zhang2003Sci,Lv2003EPJA,Zhang2005NPA,Meng2006PPNP}.
The first mass table including continuum for nuclei with $8\leq Z \leq 120$ has recently been constructed 
with the RCHB theory \cite{Xia2018ADNDT}.

Since most nuclei are deformed, Zhou and his collaborators have developed the deformed relativistic 
Hartree-Bogoliubov theory in continuum (DRHBc) to unifiedly describe the effects of the pairing correlation, 
continuum and deformation \cite{Zhou2010PRC,Li2012PRC}. The DRHBc theory has been applied to study the 
halo phenomena in magnesium isotopes and predicted an interesting shape decoupling between the core and 
the halo \cite{Zhou2010PRC,Li2012PRC}. Recently it was used to resolve the puzzles concerning the radius 
and configuration of valence neutrons in $^{22}$C \cite{Sun2018PLB}, and investigate the particles in the 
classically forbidden regions \cite{Zhang2019PRC}. The DRHBc theory was also extended to incorporate the 
blocking effect in odd-nucleon systems \cite{Li2012CPL}, and the density-dependent meson-nucleon couplings \cite{Chen2012PRC}.

In the DRHBc theory, the relativistic Hartree-Bogoliubov (RHB) equation is solved in a Dirac 
Woods-Saxon basis \cite{Zhou2003PRC}. In addition, for axially deformed nuclei, the potentials 
and densities are expressed in terms of the multipole expansion with Legendre polynomials.
In the existing literature, numerical checks for the DRHBc calculations, for instance, the 
convergence of box size and basis energy cutoff, have been carefully performed 
\cite{Zhou2003PRC,Li2012PRC,Zhang2019DRHBc}. In Ref.~\refcite{Li2012PRC}, the truncation for the 
Legendre expansion order was chosen as 4 for light nuclei. However, up to now there is no systematic 
investigation for the dependence on the Legendre expansion of the DRHBc calculations.
Since nuclei may differ largely in both mass number and deformation, it is necessary to examine the 
convergence of the Legendre expansion for nuclei in these different cases, and find out how the DRHBc 
solutions depend on the high-order terms.

In this paper, a light nucleus $^{20}$Ne and a heavy nucleus $^{242}$U are calculated as examples to 
investigate the convergence of the multipole expansion with Legendre polynomials in the DRHBc theory.
To study the dependence on the deformation, the constrained DRHBc calculations are performed.
In Sec.~\ref{sec:th}, we give a brief theoretical framework of the DRHBc theory.
The numerical details used in the calculations are given in Sec.~\ref{sec:num}.
In Sec.~\ref{sec:res} we present our results of the convergence check for the DRHBc.
Finally, the work is summarized in Sec.~\ref{sec:sum}.

\section{Theoretical framework}
\label{sec:th}

The details of the DRHBc theory can be found in Refs.~\refcite{Zhou2010PRC,Li2012PRC} with non-linear 
meson exchange effective interaction, in Ref.~\refcite{Chen2012PRC} with density-dependent meson-nucleon 
couplings, and in Ref.~\refcite{Zhang2019DRHBc} with point-coupling interaction.
Here we only present briefly the formalism with the point-coupling interaction.
By using the Bogoliubov transformation to include the pairing correlation, one can derive the RHB equation 
for nucleons \cite{Kucharek1991ZPA}
\begin{equation}
	\label{eq:RHB}
	\begin{pmatrix}
	h_D - \lambda_\tau & \Delta \\
	-\Delta^* & -h_D + \lambda_\tau
	\end{pmatrix}
	\begin{pmatrix} U_k \\ V_k \end{pmatrix}
	= E_k \begin{pmatrix} U_k \\ V_k \end{pmatrix},
\end{equation}
where $\lambda_\tau ~ (\tau=n,~p)$ is the chemical potential of neutron or proton, $E_k$ and $(U_k,V_k)^T$ 
are the quasiparticle energy and wave function, and $h_D$ is the Dirac Hamiltonian
\begin{equation}
	\label{eq:hD}
	h_D(\bm{r}) = \bm{\alpha} \cdot \bm{p} + V(\bm{r}) + \beta [M + S(\bm{r})],
\end{equation}
with the scalar and vector potentials $S(\bm{r})$ and $V(\bm{r})$,
\begin{align}
	\label{eq:S}
	S(\bm{r}) & = \alpha_S \rho_S + \beta_S \rho_S^2 + \gamma_S \rho_S^3 + \delta_S \Delta\rho_S, \\
	\label{eq:V}
	V(\bm{r}) & = \alpha_V \rho_V + \gamma_V \rho_V^3 + \delta_V \Delta\rho_V + eA_0 + \alpha_{TV} \tau_3 \rho_{TV} 
                + \delta_{TV} \Delta \tau_3 \rho_{TV}.
\end{align}
Here $M$ is the nucleon mass, and $\alpha_S, ~ \alpha_V, ~ \alpha_{TV}, ~ \beta_S, ~ \gamma_S, ~ \gamma_V, 
~ \delta_S, ~ \delta_V$ and $\delta_{TV}$ are the coupling constants; $A_0$ is the Coulomb field, and $\rho_S, 
~ \rho_V$ and $\rho_{TV}$ refer to the densities in scalar, vector, and isovector-vector channels, respectively \cite{Zhao2012PRCmass}.
The pairing potential is
\begin{align}
	\begin{aligned}
	\label{eq:Delta}
	\Delta(\bm{r}_1s_1p_1,\bm{r}_2s_2p_2)
	= \sum_{s_1'p_1'}\sum_{s_2'p_2'} V^{pp}(\bm{r}_1,\bm{r}_2;s_1p_1,s_2p_2;s_1'p_1',s_2'p_2') \times \kappa (\bm{r}_1s_1'p_1',\bm{r}_2s_2'p_2'),
	\end{aligned}
\end{align}
where $s$ represents the spin degree of freedom, $p$ represents the large or small component in Dirac spinor, 
$V^{pp}$ is the pairing force, and $\kappa$ is the pairing tensor \cite{Ring1980NMBP}.

The iterative solution of these RHB equations yields the quasiparticle levels and expectation values of 
total energy, quadrupole moments, etc. The total energy of a nucleus is \cite{Xia2018ADNDT}
\begin{align}
	\begin{aligned}[b]
	E_{\mathrm{RHB}} =
	& \sum_{k} (\lambda_\tau-E_{k}) v_k^2 - E_{\mathrm{pair}} \\
	& - \int \mathrm{d}^3 \bm{r} \bigg [ \frac{1}{2}\alpha_{S}\rho_S^2 + \frac{1}{2}\alpha_{V}\rho_{V}^2 
      + \frac{1}{2}\alpha_{TV}(\rho_{TV})^2 \\
	& + \frac{2}{3}\beta_{S}\rho_{S}^3 + \frac{3}{4}\gamma_{S}\rho_{S}^4 + \frac{3}{4}\gamma_{V}(\rho_{V})^4 
      + \frac{1}{2}\delta_{S}\rho_S\Delta\rho_S \\
	& + \frac{1}{2}\delta_{V}\rho_{V}\Delta\rho_{V}
	+\frac{1}{2}\delta_{TV}\rho_{TV}\Delta\rho_{TV} + \frac{1}{2}eA_{0}\rho_{p} \bigg ] \\
	& + E_{\mathrm{c.m.}},
	\end{aligned}
\end{align}
where
\begin{equation}
	v_k^2 = \int \mathrm{d}^3 \bm{r} V_k^\dagger(\bm{r}) V_k(\bm{r}) ,
\end{equation}
$E_{\mathrm{pair}}$ is the pairing energy, and $E_{\mathrm{c.m.}}$ is the microscopic center-of-mass 
correction energy \cite{Bender2000EPJA,Long2004PRC,Zhao2009CPL}. The intrinsic quadrupole moment is calculated by
\begin{equation}
	Q_{\tau,2} = \sqrt{\frac{16\pi}{5}} \langle {r^2 Y_{20}(\theta,\phi)} \rangle .
\end{equation}
The quadrupole deformation parameter is obtained from the quadrupole moment by
\begin{equation}
\beta_{\tau,2} = \frac{ \sqrt{5 \pi} Q_{\tau,2} } {3 N_\tau \langle r_\tau^2\rangle} \ ,
\end{equation}
where $N_\tau$ refers to the number of neutron, proton, or nucleon.

In the DRHBc theory, the RHB equation \eqref{eq:RHB} is solved in a Dirac Woods-Saxon basis \cite{Zhou2003PRC}.
For the axially deformed potentials and densities with spatial reflection symmetry in Eqs.~\eqref{eq:hD}, 
\eqref{eq:S}, \eqref{eq:V} and \eqref{eq:Delta}, it is convenient to express the angular dependence in terms 
of the multipole expansion, where the expansion basis functions are Legendre polynomials \cite{Price1987PRC}
\begin{equation}
	f(\bm{r}) = \sum_\lambda f_\lambda(r)  \mathrm{P}_\lambda(\cos\theta),
	\qquad \lambda = 0, ~ 2, ~ 4, ~ \cdots, ~ \lambda_{\max},
\end{equation}
with
\begin{equation}
	f_\lambda(r) = \frac{2\lambda+1}{4\pi} \int \mathrm{d}\Omega f(\bm{r}) \mathrm{P}_\lambda(\Omega),
\end{equation}
where $f(\bm{r})$ refers to these potentials or densities. In practical numerical implementations, 
a truncation for the expansion order, $\lambda_{\max}$, has to be introduced.

\section{Numerical details}
\label{sec:num}

To investigate the convergence of the multipole expansion with Legendre polynomials in the DRHBc theory, 
the even-even nuclei $^{20}$Ne and $^{242}$U are calculated as examples. The density functional adopted 
is PC-PK1 \cite{Zhao2010PRC}, which has turned out to provide one of the best density functional descriptions 
for nuclear masses so far \cite{Zhao2012PRCmass,Zhang2014FoP,Lu2015PRC}. In the present convergence check 
against the Legendre expansion truncation $\lambda_{\max}$, other numerical parameters are fixed in the DRHBc calculations.
The box size is fixed at $R_{\max}=20~\mathrm{fm}$, and the mesh size $\Delta r=0.1~\mathrm{fm}$.
For the Woods-Saxon basis space, the angular momentum cutoff is $J_{\max}=23/2~\hbar$, and the energy cutoff 
is $E_{\mathrm{cut}}^+=300~\mathrm{MeV}$ for positive-energy states in the Fermi sea.
The number of negative-energy states in the Dirac sea is set the same as that of positive-energy states in the Fermi sea.
These numerical conditions above have been examined to converge well in Refs.~\refcite{Zhou2003PRC,Li2012PRC,Zhang2019DRHBc}.
When the pairing correlation is taken into account, we use the density-dependent zero-range force \cite{Meng1998NPA,Li2012PRC}, 
where the pairing strength is taken as $V_0=-325~\mathrm{MeV~fm^3}$, and a sharp cutoff 
$E_{\mathrm{cut}}^{\mathrm{q.p.}}=100~\mathrm{MeV}$ in the quasiparticle space is adopted \cite{Zhang2019DRHBc}.

\section{Results and discussion}
\label{sec:res}

\begin{figure}[htbp]
  \centering
  \includegraphics[width=0.9\linewidth]{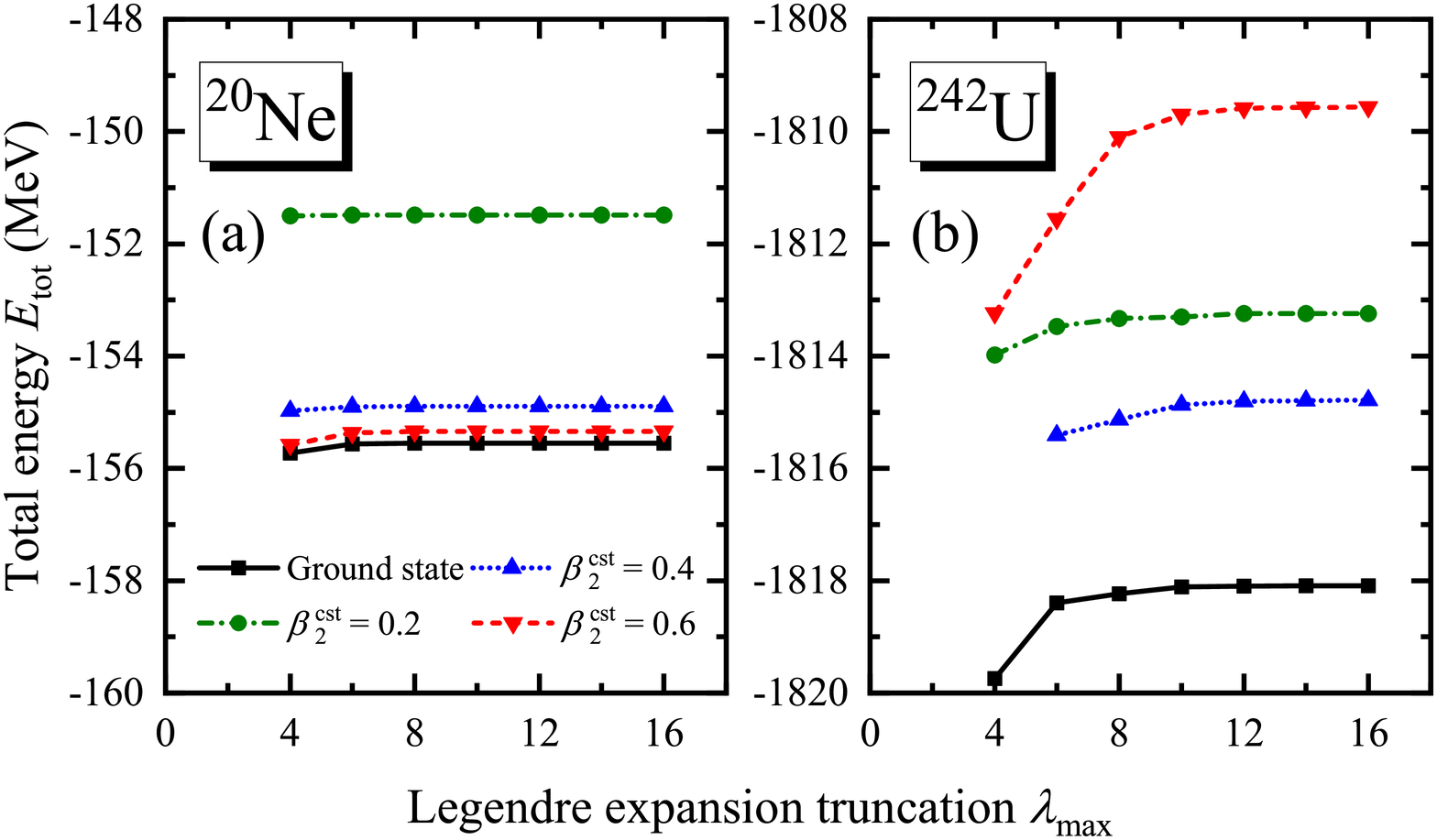}
  \caption{ Total energies of $^{20}$Ne (a) and $^{242}$U (b) as functions of the Legendre expansion 
  truncation $\lambda_{\max}$ in the DRHBc calculations with PC-PK1.
  The black solid curve represents the results in the ground state, and the green dotted-dashed, 
  blue dotted, and red dashed curves display the results with deformations constrained at 
  $\beta_2^{\mathrm{cst}}=0.2$, $0.4$, and $0.6$, respectively.
  Here the pairing correlation is not considered.
  }
  \label{fig:Etot}
\end{figure}

In Fig.~\ref{fig:Etot}, the total energies of a light nucleus $^{20}$Ne and a heavy nucleus $^{242}$U, 
calculated by using the DRHBc with PC-PK1, in the ground state and with deformation constrained at 
$\beta_2^{\mathrm{cst}}=0.2$, 0.4 and 0.6, are plotted as functions of the Legendre expansion truncation $\lambda_{\max}$.
As a first step, in order to avoid possible coupling effects from the scattering of Cooper pairs, 
the pairing correlation is not considered in the calculations. As seen in Fig.~\ref{fig:Etot}, 
apparently, when $\lambda_{\max}$ increases, the total energies converge well in all cases considered here.
For example, the deviation of the ground-state energy of $^{20}$Ne is $0.178~\mathrm{MeV}$ with 
$\lambda_{\max}=4$ from that with $\lambda_{\max}=16$, and it is $0.018~\mathrm{MeV}$ with 
$\lambda_{\max}=6$, only about $0.01\%$ of the total energy. In each panel, the ground state corresponds to 
the lowest energy, and converges to $-155.548~\mathrm{MeV}$ and $-1818.093~\mathrm{MeV}$ for $^{20}$Ne and 
$^{242}$U, respectively. It is noted that, as the ground state is obtained from the unconstrained calculation, 
the deformation also changes with $\lambda_{\max}$. The quadrupole deformations of $^{20}$Ne and $^{242}$U 
with $\lambda_{\max}=4$ are $\beta_2=0.55$ and $0.32$, respectively, which converge to $\beta_2=0.54$ and 
$0.31$ with $\lambda_{\max}\geq 6$. Therefore, it is shown that both the total energy and deformation of 
the ground state converge well with $\lambda_{\max}$.

To study the influence of the deformation on the convergence of $\lambda_{\max}$, the results calculated 
at different constrained deformations are compared. In Fig.~\ref{fig:Etot}, the deviation of the total energy 
of $^{20}$Ne calculated with $\lambda_{\max}=4$ from that with $\lambda_{\max}=16$ at $\beta_2^{\mathrm{cst}}=0.2$ 
is $0.016$~MeV, while it is $0.081$~MeV at $\beta_2^{\mathrm{cst}}=0.4$, and $0.238$~MeV at $\beta_2^{\mathrm{cst}}=0.6$.
Similarly, the deviation of the total energy of $^{242}$U also increases with the constrained deformation.
This shows that the dependence of the total energy on $\lambda_{\max}$ increases with deformation.

Then to figure out the influence of the nuclear mass on the convergence of $\lambda_{\max}$, we take 
the largest deformation considered here, i.e. $\beta_2^{\mathrm{cst}}=0.6$, which poses a higher requirement 
for $\lambda_{\max}$ than others. For the light nucleus $^{20}$Ne shown in Fig.~\ref{fig:Etot}(a), the 
deviations of the total energies with $\lambda_{\max}=4$ and 6 from that with $\lambda_{\max}=16$ are 
$0.238$~MeV and $0.029$~MeV, which are about $0.15\%$ and $0.02\%$ of the total energies, respectively, 
marking a good convergence. However, for the heavy nucleus $^{242}$U shown in Fig.~\ref{fig:Etot}(b), 
the difference between the total energies with $\lambda_{\max}=6$ and 16 is $1.986$~MeV, which is a 
relatively large value. By increasing $\lambda_{\max}$ to 8 and 10, the deviations of the total energies 
from that with $\lambda_{\max}=16$ are lowered to $0.534$~MeV and $0.135$~MeV, which are about $0.03\%$ 
and $0.01\%$ of the total energies, respectively. Therefore, to reach a same accuracy of the total energy, 
even to a same relative accuracy by percent, a larger $\lambda_{\max}$ is required for a heavy nucleus than a light one.

\begin{figure}[htbp]
  \centering
  \includegraphics[width=0.8\linewidth]{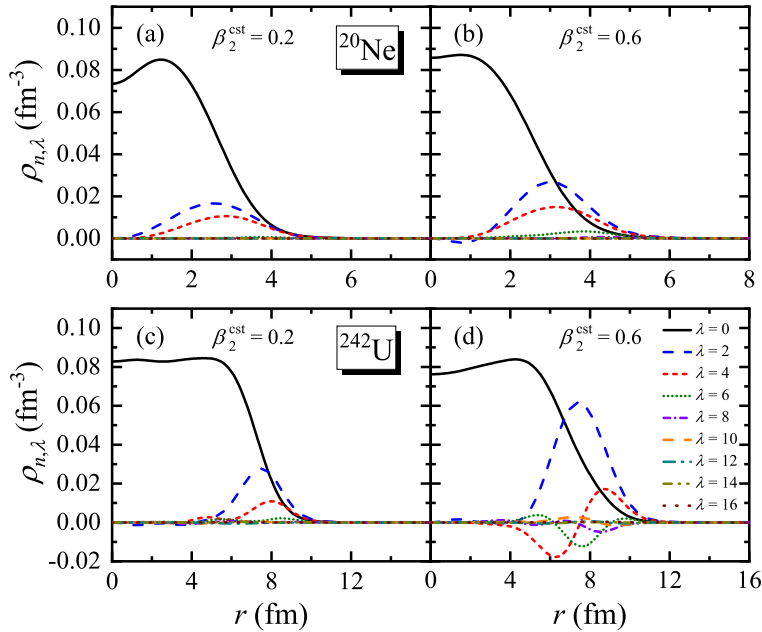}
  \caption{ Decompositions into different Legendre expansion components of the neutron densities of 
  $^{20}$Ne with deformation constrained at $\beta_2=0.2$ (a) and 0.6 (b), and of $^{242}$U at 
  $\beta_2=0.2$ (c) and 0.6 (d), in the DRHBc calculations with $\lambda_{\max}=16$.
  }
  \label{fig:dens}
\end{figure}

In order to intuitively understand the behaviors of the convergence of $\lambda_{\max}$ shown in Fig.~\ref{fig:Etot}, 
the neutron densities $\rho_n$ of $^{20}$Ne and $^{242}$U in the DRHBc calculations with $\lambda_{\max}=16$, are 
both decomposed into different Legendre expansion components $\rho_{n,\lambda}$, as shown in Fig.~\ref{fig:dens}.
For each nucleus, the quadrupole deformation is constrained at two values, $\beta_2^{\mathrm{cst}}=0.2$ and $0.6$.
It can be seen that in each panel of Fig.~\ref{fig:dens}, the $\lambda=0$ component $\rho_{n,0}$ is always the most 
important one, and with the increase of $\lambda$, the corresponding component $\rho_{n,\lambda}$ becomes smaller.

For the light nucleus $^{20}$Ne, at $\beta_2^{\mathrm{cst}}=0.2$, as shown in Fig.~\ref{fig:dens}(a), the neutron 
density components of $\lambda=0 \sim 4$ are clearly seen; while at $\beta_2^{\mathrm{cst}}=0.6$ as shown in 
Fig.~\ref{fig:dens}(b), it is seen that the components of $\lambda=0\sim 6$ are obvious, where the nonzero-$\lambda$ 
ones are relatively larger than those at $\beta_2^{\mathrm{cst}}=0.2$. Therefore, with the increase of deformation, 
the high-order components get larger, and this is consistent with the fact that the dependence of the total energy 
on $\lambda_{\max}$ increases with deformation. In addition, at $\beta_2^{\mathrm{cst}}=0.6$ the components of 
$\lambda>6$ almost vanish, and this is consistent with the fact in Fig.~\ref{fig:Etot}(a) that the total energy 
of $^{20}$Ne gets well converged with $\lambda_{\max}=6$. For the heavy nucleus $^{242}$U, at $\beta_2^{\mathrm{cst}}=0.2$, 
as shown in Fig.~\ref{fig:dens}(c), we can clearly see the components of $\lambda=0 \sim 4$, and also $\lambda=6$ 
with small values; while at $\beta_2^{\mathrm{cst}}=0.6$ as shown in Fig.~\ref{fig:dens}(d), the relatively 
obvious components are those of $\lambda=0\sim 8$, and $\lambda=10$ with small values as well.
This is consistent with the convergence of the total energy of $^{242}$U in Fig.~\ref{fig:Etot}(b).
Therefore, comparing Fig.~\ref{fig:dens}(a) with (c), and (b) with (d), it is found at the same constrained 
deformation, the high-order components of a heavy nucleus play a more important role than that of a light one, 
which also corresponds to the influence of nuclear mass on the convergence of $\lambda_{\max}$ obtained from Fig.~\ref{fig:Etot}.

\begin{figure}[htbp]
  \centering
  \includegraphics[width=0.8\linewidth]{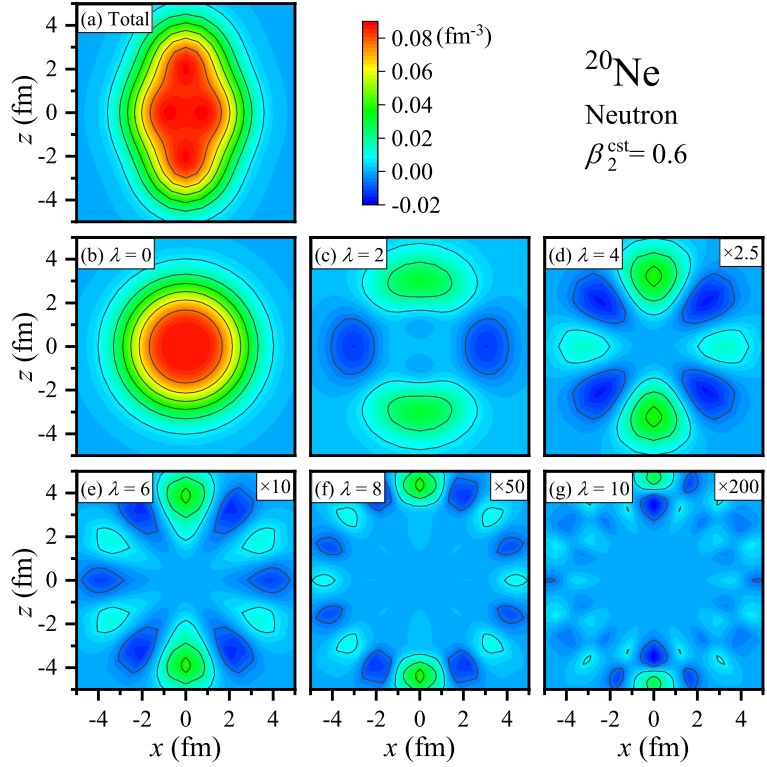}
  \caption{ Neutron density distributions in $x$-$z$ plane of $^{20}$Ne with deformation constrained at 
  $\beta_2^{\mathrm{cst}}=0.6$ in the DRHBc calculations with $\lambda_{\max}=16$: (a) the total neutron 
  density distribution, and (b) to (g) the decompositions into different Legendre expansion components of 
  neutron density distribution, respectively. The number labeled on the upper right corner of each panel 
  from (d) to (g) is the multiplying factor for the corresponding component.
  }
  \label{fig:Ne20_ndensxz}
\end{figure}

To see more clearly and visually the spatial distributions of the different components in the Legendre expansion, 
in Fig.~\ref{fig:Ne20_ndensxz} the total and decomposed neutron density distributions of $^{20}$Ne with deformation 
constrained at $\beta_2^{\mathrm{cst}}=0.6$ in the DRHBc calculations with $\lambda_{\max}=16$, is plotted in $x$-$z$ plane.
Fig.~\ref{fig:Ne20_ndensxz}(a) gives the total neutron density distribution, and the following six panels, i.e., (b) to (g), 
show the decompositions into different Legendre expansion components of neutron density distribution, respectively.
To see the spatial distribution of high-order components more clearly, each component in panels (d) to (g) is multiplied 
by a factor, which can provide a reference for the order of magnitude of the corresponding density component.
For $\lambda=4$, 6, 8 and 10, the multiplying factor is 2.5, 10, 50, and 200, showing that the corresponding component 
decreased by half a magnitude one-by-one. This is consistent with the fact in Fig.~\ref{fig:dens} that a higher $\lambda$ 
corresponds to a smaller component $\rho_{n,\lambda}$. Furthermore, it can be seen in Fig.~\ref{fig:Ne20_ndensxz}(a) that 
the shape of the total neutron density distribution of $^{20}$Ne at $\beta_2^{\mathrm{cst}}=0.6$ is nearly prolate;
in panel (b) the $\lambda=0$ component is exactly spherically symmetric; and in panels (c) to (g) the components of 
different $\lambda\neq 0$ have the corresponding shapes with the angular part satisfying spherical harmonics $\mathrm{Y}_{\lambda 0}$.
In fact, the summation of the $\lambda=0$, 2 and 4 components have already provided a relatively accurate description 
to the total density distribution in Fig. \ref{fig:Ne20_ndensxz}(a), while other components with higher $\lambda$ provide 
higher-order corrections to the shape.

\begin{figure}[htbp]
  \centering
  \includegraphics[width=0.6\linewidth]{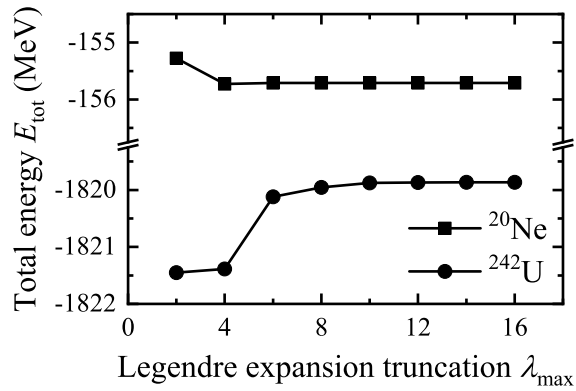}
  \caption{ Ground-state energies of $^{20}$Ne and $^{242}$U as functions of the Legendre expansion 
  truncation $\lambda_{\max}$ in the DRHBc calculations with PC-PK1.
  Here the pairing correlation is included.
  }
  \label{fig:Etot_pair}
\end{figure}

In the discussions above, the pairing correlation is neglected to avoid possible coupling effects from 
the scattering of Cooper pairs. However, in open shell nuclei, the pairing correlation plays an important 
role \cite{Ring1980NMBP}. Therefore, to make it an end, it is necessary to study the convergence of the 
Legendre expansion truncation $\lambda_{\max}$ with the pairing correlation included.
Figure~\ref{fig:Etot_pair} shows the ground-state energies of $^{20}$Ne and $^{242}$U as functions of 
$\lambda_{\max}$, respectively. For $^{20}$Ne the calculated ground-state energy converges to 
$E_{\mathrm{tot}}=-155.708~\mathrm{MeV}$ with deformation $\beta_2=0.48$; for $^{242}$U the corresponding 
results are $E_{\mathrm{tot}}=-1819.865~\mathrm{MeV}$ with $\beta_2=0.30$. With the pairing correlation included, 
the total energy $E_{\mathrm{tot}}$ becomes lower and deformation $\beta_2$ smaller, as usually expected due to the pairing effects.

More explicitly, for $^{20}$Ne the deviations of total energies with $\lambda_{\max}=4$ and 6 from 
that with $\lambda_{\max}=16$ are about $0.018$~MeV and $0.002$~MeV, respectively, which are about 
$0.012\%$ and $0.002\%$ of the total energies, whereas for $^{242}$U the deviations of total energies 
with $\lambda_{\max}=4$, $6$ and $8$ from that with $\lambda_{\max}=16$ are about $1.521$~MeV, $0.255$~MeV, 
and $0.091$~MeV, which are about $0.084\%$, $0.014\%$, and $0.005\%$ of the total energies, respectively.
Therefore, the total energy converges well, and to reach a same relative accuracy by percent, a heavy 
nucleus requires a higher $\lambda_{\max}$ than a light one. This is similar to the results in 
Fig.~\ref{fig:Etot}, which means the inclusion of the pairing correlation does not change the conclusions 
from the case without the pairing correlation.

\section{Summary}
\label{sec:sum}

In summary, a light nucleus $^{20}$Ne and a heavy nucleus $^{242}$U have been calculated to investigate 
the convergence of the multipole expansion with Legendre polynomials in the deformed relativistic 
Hartree-Bogoliubov theory in continuum. The total energy converges well with the expansion truncation 
$\lambda_{\max}$ both in the absence of and presence of the pairing correlation, either in the ground state 
or at a constrained quadrupole deformation. It is interesting to find that to reach a same accuracy of 
the total energy, even to a same relative accuracy by percent, a larger $\lambda_{\max}$ is required 
for a heavy nucleus than a light one. The dependence of the total energy on $\lambda_{\max}$ increases 
with deformation. Furthermore, by decompositions of the neutron density distribution, it is shown that 
its $\lambda=0$ component is exactly spherically symmetric; and its $\lambda\neq 0$ components have 
the corresponding shapes with the angular part satisfying spherical harmonics $\mathrm{Y}_{\lambda 0}$.
A higher-$\lambda$ component has a smaller contribution to the density distribution, and with the increase 
of deformation, the high-order components get larger. It is also shown at the same constrained deformation, 
the high-order components of a heavy nucleus play a more important role than those of a light one.

\section*{Acknowledgements}

	The authors would like to express gratitude to J. Meng for constructive guidance and valuable suggestions, 
    and to J. H. Chi, Z. X. Ren, Y. K. Wang and P. W. Zhao for useful suggestions.
	The discussions with all members of the DRHBc mass table collaboration are highly acknowledged, in particular 
    during ``the 2nd workshop on nuclear mass table with DRHBc theory''.
	This work was partly supported by the National Science Foundation of China (NSFC) under Grants No. 11875075, 
    No. 11621131001, No. 11935003, and No. 11975031 and the National Key R\&D Program of China 
    (Contracts No. 2018YFA0404400 and No. 2017YFE0116700).
	

\end{document}